\documentclass[11pt]{article}

\usepackage{graphicx}
\usepackage{epstopdf}
\usepackage{natbib}
\usepackage{xspace}
\usepackage{xcolor}
\usepackage{amsmath, amssymb}

\usepackage{geometry}
\usepackage{subcaption}

\usepackage{siunitx}

\newcommand\etc{\textit{etc.}\xspace}
\newcommand\eg{\textit{e.g.}\xspace}
\newcommand\ie{\textit{i.e.}\xspace}

\renewcommand\Re{\mbox{$\mathrm{Re}$}}
\renewcommand\Im{\mbox{$\mathrm{Im}$}}

\def \B {\mbox{${\boldsymbol B}$}}

\def \J {\mbox{${\boldsymbol j}$}}

\def \xpc {\bar{x}}
\def \ypc {\bar{y}}

\def \Gb {\bar{G}}
\def \Ib {\bar{I}}
\def \Kb {\bar{K}}

\def \Rh {\hat{\mathbf R}}
\def \phih {\hat{\boldsymbol{\phi}}}
\def \zh {\hat{\boldsymbol z}}
\providecommand\bnabla{\boldsymbol{\nabla}}

\def \eb {\mathbf{e}}

\def \Jac {J}
\def \Jb {\bar{J}}
\def \pb {\bar{p}}

\usepackage{unicode}
\newcommand{\csso}{\fontencoding{LECO}\selectfont\char215}
\newcommand{\iotaslash}{\hspace*{0.1em}\iota\hspace*{-0.45em}\text{\csso}}

\newcommand{\poiss}[2]{\left\{ #1, #2 \right\}}

\providecommand\bnabla{\boldsymbol{\nabla}}

\providecommand\bphi{\boldsymbol{\phi}}

\usepackage{authblk}

\title{Perturbing an axisymmetric magnetic equilibrium to obtain a quasi-axisymmetric stellarator}
\author[1]{G.G. Plunk\thanks{gplunk@ipp.mpg.de}}
\affil[1]{Max Planck Institute for Plasma Physics, Greifswald 17491, Germany}

\begin{document}
\maketitle

\begin{abstract}
It is demonstrated that finite-pressure, approximately quasi-axisymmetric stellarator equilibria can be directly constructed (without numerical optimization) via perturbations of given axisymmetric equilibria.  The size of such perturbations is measured in two ways, via the fractional external rotation and, alternatively, via the relative magnetic field strength, \ie the average size of the perturbed magnetic field, divided by the unperturbed field strength.  It is found that significant fractional external rotational transform can be generated by quasi-axisymmetric perturbations, with a similar value of the relative field strength, despite the fact that the former scales more weakly with the perturbation size.  High mode number perturbations are identified as a candidate for generating such transform with local current distributions.  Implications for the development of a general non-perturbative solver for optimal stellarator equilibria is discussed.
\end{abstract}

\section{Introduction}

Quasi-symmetry is a property of magnetic fields that ensures the confinement of collisionless particle orbits.  Axisymmetric magnetic equilibria possess this property in a trivial sense, whereas the related class of stellarators, called quasi-axisymmetric (QAS), satisfy the symmetry in a way that is hidden to the naked eye \citep{nuehrenberg-QAS}.

The close relationship between axisymmetry and QAS suggests that the second class may be continuously connected to the first, and in particular that QAS stellarators may be obtained by deformation of axisymmetric equilibria \citep{boozer-ITER}.  It has also been suggested that modifying tokamak equilibria by non-axisymmetric shaping might help overcome the stability issues that plague them, and a previous study, using conventional numerical optimization, has demonstrated that suitable QAS may indeed be found as deformed tokamak equilibria \citep{ku-boozer}.  The idea of passively stabilizing a tokamak by non-axisymmetric perturbations is also supported by a number of experimental results, when the perturbation generates a sufficient ``external'' boost in rotational transform \citep{w7a, cth}.

Solving the MHD equilibrium problem for optimal stellarator equilibria, without the use of numerical optimization algorithms (\ie ``direct construction'' of optimal solutions), is potentially beneficial due to the speedup offered \citep{landreman_sengupta_plunk_2019}.  So far, the only ways to do this have involved approximations to the problem like small distance from the magnetic axis \citep{garren-boozer-1, garren-boozer-2, landreman_sengupta_2018, landreman_sengupta_plunk_2019, plunk_landreman_helander_2019}, or small deviation from axi-symmetry \citep{plunk-helander-qas}.  Solving these approximate problems can also lead to fundamental insights about the properties of equilibria, and the size of the solution space.

There are possible practical advantages of directly constructing QAS stellarator equilibria via perturbation of axisymmetric equilibria, as compared to conventional optimization.  For instance, the general perturbation can be constructed as a sum of independent QAS modes with different toroidal mode numbers.  After pre-computation of these modes, the corresponding space of QAS equilibria can be easily scanned, without further computational cost, whereas each step of a conventional optimizer involves solving the equilibrium problem anew.  Also, there is no fundamental constraint on axisymmetric equilibrium measures, like aspect ratio, so these may be set arbitrarily to explore new areas of stellarator design space, which may have been inaccessible with conventional optimization.

In a previous paper \citep{plunk-helander-qas}, it was proved that nearly axi-symmetric magnetic fields can be constructed to satisfy the condition of quasi-axisymmetry on a single magnetic surface.  These solutions, however, apply only to vacuum conditions, where the plasma itself does not contribute significantly to the magnetic field.  The present work considers the more general case of finite pressure equilibria.  Formidable challenges are present in this general problem, starting with an increased complexity arising from the nonlinear coupling of multiple fields.  The presence of singularities in the force balance equation makes the general problem of obtaining equilibria ill-posed, even without the requirement of satisfying a special symmetry.  As we will show, the issue of force balance singularities may be overcome, at least at first order in the expansion, by suitable choice of the zeroth-order rotational transform profile.  The complexity of the system, however, makes it more difficult to establish existence of solutions by the same methods employed by \citet{plunk-helander-qas}.  We therefore turn to devising a method to numerically solve the system.  This, as we find, gives evidence that the same problem as solved in the vacuum limit by \citet{plunk-helander-qas}, namely the problem of finding a perturbation of specified toroidal mode number $N$ that satisfies the condition of QAS on a single magnetic surface, is indeed well-posed, at least in some practical sense.

The contents of the paper are as follows.  In section \ref{preliminaries-sec} the basic equations and notation are established, and the ``inverse'' MHD equilibrium problem formulation is described.  In section \ref{expansion-sec}, the expansion about axi-symmetry is performed, and the equations are given to find perturbations satisfying QAS on a specified magnetic surface.  The issue of force balance singularities is discussed, and a strategy to overcome them is described.  In section \ref{numerical-sec}, a numerical method is described to solve the first order system, and a set of solutions are given, based on a zeroth order ITER-like equilibrium.  The VMEC \citep{Hirshman} and BOOZ\_XFORM \citep{booz-xform} codes are used to demonstrate that the solutions can satisfy the appropriate level of QAS as predicted by the theory.

\section{Preliminaries}\label{preliminaries-sec}

The MHD equilibrium equations are

\begin{eqnarray}
&\bnabla\times\B = \mu_0\J,\label{ampere-eqn}\\
&\bnabla\cdot\B = 0.\label{div-B-eqn}\\
&\J\times\B = \bnabla \psi \frac{d p}{d\psi},\label{force-balance-eqn}
\end{eqnarray}

\noindent We assume topologically toroidal magnetic surfaces, here labeled by the flux function $\psi$.  To solve these equations, we use a similar approach as previous works \citet{garren-boozer-1, garren-boozer-2, hegna-2000, boozer-2002, Weitzner}.  Boozer angles are denoted $\theta$ and $\varphi$.  The contravariant form of $\B$ is written

\begin{equation}
\B_{\mathrm{con}} = \bnabla\psi \times \bnabla\theta - \iotaslash(\psi) \bnabla \psi \times \bnabla \varphi,\label{B-contra-eqn}
\end{equation}

\noindent where $\iotaslash$ is the rotational transform, and $2\pi \psi$ is the toroidal flux.  This form of $\B$ satisfies zero divergence, assuming flux-surface geometry.  The covariant form is written

\begin{equation}
\B_{\mathrm{cov}} = G(\psi) \bnabla\varphi + I(\psi) \bnabla\theta + K(\psi, \theta, \varphi) \bnabla \psi.\label{B-covar-eqn}
\end{equation}
This form is a consequence of $\J \cdot \bnabla \psi = 0$ (\ref{force-balance-eqn}), and Ampere's law (\ref{ampere-eqn}); see \eg \cite{helander-review}.  

The basic strategy to find an equilibrium is to assert $\B_{\mathrm{con}} = \B_{\mathrm{cov}}$ together with force balance (\ref{force-balance-eqn}), relying on the fact that these forms of the magnetic field incorporate Eqns.~\ref{ampere-eqn} and \ref{div-B-eqn} as well as the assumption of topologically toroidal magnetic surfaces.  Either the magnetic coordinates $\psi$, $\theta$, and $\varphi$ can be considered as the unknown functions of spatial coordinates (``direct formulation"), or the coordinate mapping ${\bf x}(\psi, \theta, \varphi)$ can be considered as the unknown function of magnetic coordinates (``inverse formulation'').   Both formulations are used here.  

It is convenient at zeroth order to solve the Grad-Shafranov equation (\eg using the direct formulation).  This means that we are able to specify the axisymmetric shape of the outer magnetic surface.  We will also specify the current and pressure profiles at this stage, and consider them as fixed for the remainder of the calculation.  We will use the indirect formulation for the problem at next order, \ie the problem of QAS-preserving perturbations, as it casts the problem as a fixed boundary problem with QAS as the boundary condition.

\subsection{Problem formulation}

With the inverse formulation, the independent variables of the problem are the magnetic coordinates, and QAS is expressed as a simple constraint, $\partial B/\partial \varphi = 0$.  Instead of using magnetic flux as a coordinate, we will use a coordinate system based on a dimensionless radial coordinate $\rho = \sqrt{\psi/\psi_b}$, where $\psi_b$ denotes the value of $\psi$ on the boundary surface.  Note that most physical quantities are not analytic in $\psi$ at the magnetic axis ($\psi = 0$), but can be expanded in $\rho$ \citep{garren-boozer-1}.  This idea is motivated by considering $\rho$ and $\theta$ as polar coordinates and then assuming that a Taylor expansion can be made in the pseudo-cartesian coordinates $
\xpc = \rho \cos(\theta)$ and $\ypc = \rho \sin(\theta)$.

With the inverse formulation, the unknown of the theory is the coordinate mapping ${\bf x}(\rho, \theta, \varphi)$, and the equilibrium equations are written in terms of various derivatives $\partial {\bf x}/\partial \rho$, and so forth.  These equations can be translated into equations involving the metrics via the usual identities (reviewed in Appx.~\ref{geometry-appx}).  The equation $\B_{\mathrm{con}} = \B_{\mathrm{cov}}$ becomes

\begin{equation}
\rho\left(\frac{\partial {\bf x}}{\partial \varphi} + \iotaslash \frac{\partial {\bf x}}{\partial \theta}\right) = \Gb \frac{\partial {\bf x}}{\partial \rho}\times\frac{\partial {\bf x}}{\partial \theta} + \Ib \frac{\partial {\bf x}}{\partial \varphi}\times\frac{\partial {\bf x}}{\partial \rho} + \Kb \frac{\partial {\bf x}}{\partial \theta}\times\frac{\partial {\bf x}}{\partial \varphi}.\label{the-MHD-eqn}
\end{equation}
Force balance can be expressed as a scalar equation, since it only has a component in the $\bnabla \rho$ direction.  One uses $\J = \mu_0^{-1}\bnabla\times\B_{\mathrm{cov}}$ and take the scaler product of Eqn.~\ref{force-balance-eqn} with $\bnabla \theta \times \bnabla \varphi$

\begin{equation}
\left(\frac{d \Gb}{d \rho} - \frac{\partial \Kb}{\partial \varphi}\right) + \iotaslash\left(\frac{d \Ib}{d \rho} - \frac{\partial \Kb}{\partial \theta}\right) + \mu_0 \Jb \frac{d\pb}{d\rho}  = 0\label{f-balance-scaler-eqn}
\end{equation}

We introduce the following normalized quantities: $\Gb = G/(2\psi_b)$, $\Ib = I/(2\psi_b)$, $\Kb = \rho K$ and $\pb = p/(2\psi_b)^2$.  Note that we define $\Jb = \Jac/\rho$ in the limiting sense so that although $\Jac = (\bnabla \rho\cdot\bnabla\theta\times\bnabla\varphi)^{-1}$ tends to zero with $\rho$, $\Jb$ does not.  Finally, we will need the following expression for the regularized Jacobian:

\begin{equation}
\Jb = \frac{\left|\frac{\partial {\bf x}}{\partial \varphi} + \iotaslash \frac{\partial {\bf x}}{\partial \theta}\right|^2}{\Gb + \iotaslash \Ib}.\label{Jb-eqn}
\end{equation}
Defining also ${\bf \bar{B}} = {\bf B}/(2\psi_b)$ we have the useful relation $\Jb = (\Gb + \iotaslash \Ib)/\bar{B}^2$ (Eqn.~\ref{JacB-eqn-1}) so that QAS can be expressed most conveniently here as 

\begin{equation}
\frac{\partial \Jb}{\partial \varphi}  = 0.
\end{equation}
In the present work, we look for solutions that satisfy this condition on a single magnetic surface; we will not consider here the question of whether this condition might, under special circumstances, be satisfied globally, \ie uniformly in $\rho$.

\section{The expansion about axisymmetry}\label{expansion-sec}

We write the coordinate mapping ${\bf x}(\rho, \theta, \varphi)$ as a series expansion in the small parameter $\epsilon$, 

\begin{equation}
{\bf x} = {\bf x}_0 + \epsilon {\bf x}_1 + \epsilon^2 {\bf x}_2 + \dots,
\end{equation}

\noindent where ${\bf x}_0$ corresponds to the zeroth-order axisymmetric equilibrium.  We will consider the pressure $\pb$ and currents $\Gb$ and $\Ib$ as fixed to their zeroth order values (there is no loss of generality as any higher order variation in these functions can be absorbed into the zeroth order forms).  This confines our attention to axisymmetry breaking perturbations.  We must however allow the deformation to modify $\iotaslash$ and $\Kb$.

\begin{eqnarray}
\Kb(\rho, \varphi, \theta) = \Kb_0 + \epsilon \Kb_1 + \epsilon \Kb_2 + \dots,\\
\iotaslash = \iotaslash_0 + \epsilon \iotaslash_1 + \epsilon^2 \iotaslash_2 + \dots.
\end{eqnarray}

\noindent For a nearly axisymmetric equilibrium, it is sensible to take the components of Eqn.~\ref{the-MHD-eqn} along the cylindrical unit vectors $\hat{\bf R}$, $\hat{\bphi}$, $\hat{\bf z}$ (such that $\hat{\bf R} \times \hat{\bphi} = \hat{\bf z}$).

\subsection{Order $\epsilon^0$}\label{order-zero-sec}

The zeroth order coordinate mapping is (see also appendix \ref{grad-shafranov-appx})

\begin{equation}
{\bf x}_0 = \Rh R_0(\rho, \theta) + \zh Z_0(\rho, \theta),
\end{equation}
where the cylindrical unit vectors are functions of the geometric toroidal coordinate, related to the boozer angle by $\varphi = \phi + \nu$, and is expanded as

\begin{eqnarray}
\phi = \varphi - \nu_0 - \nu _1 - \dots,\label{phi-eqn}
\end{eqnarray}
so that $\phi_0 = \varphi - \nu_0$ and $\phi_1 = -\nu_1$, \etc, and, for simplicity, the unit vectors will be defined according to the zeroth order expression of the geometric toroidal angle,

\begin{eqnarray}
\Rh = \Rh(\phi_0) = \Rh(\varphi - \nu_0),\\
\phih = \phih(\phi_0) = \phih(\varphi - \nu_0).
\end{eqnarray}
With these definitions, derivatives of the zeroth order coordinate mapping are evaluated as

\begin{eqnarray}
&\frac{\partial {\bf x}_0}{\partial \rho} &= \Rh \frac{\partial R_0}{\partial \rho} + \zh \frac{\partial Z_0}{\partial \rho} - \phih R_0 \frac{\partial \nu_0}{\partial \rho} \label{e10-eq}\\
&\frac{\partial {\bf x}_0}{\partial \theta} &=  \Rh \frac{\partial R_0}{\partial \theta} + \zh \frac{\partial Z_0}{\partial \theta} - \phih R_0 \frac{\partial \nu_0}{\partial \theta} \label{e20-eq}\\
&\frac{\partial {\bf x}_0}{\partial \varphi} &= \phih R_0. \label{e30-eq}
\end{eqnarray}

The zeroth order MHD constraint is 

\begin{equation}
\rho\left(\frac{\partial {\bf x}_0}{\partial \varphi} + \iotaslash_0 \frac{\partial {\bf x}_0}{\partial \theta}\right) = \Gb \frac{\partial {\bf x}_0}{\partial \rho}\times\frac{\partial {\bf x}_0}{\partial \theta} + \Ib \frac{\partial {\bf x}_0}{\partial \varphi}\times\frac{\partial {\bf x}_0}{\partial \rho} + \Kb_0 \frac{\partial {\bf x}_0}{\partial \theta}\times\frac{\partial {\bf x}_0}{\partial \varphi}.\label{the-MHD-eqn}
\end{equation}
where, Eqns.~\ref{e10-eq}-\ref{e30-eq} can be substituted in and the equation projected along the unit vectors $\Rh$, $\phih$ and $\zh$ to obtain three coupled equations.  Note that we avoid explicitly writing the lengthy equations that result, and will do likewise with others that follow, especially when they do not give any useful insight.  Force balance is

\begin{equation}
\frac{d \Gb}{d \rho} + \iotaslash_0  \left(\frac{d \Ib}{d \rho} - \frac{\partial \Kb_0}{\partial \theta}\right) + \mu_0  \Jb_0 \frac{d\pb}{d\rho} = 0.\label{f-balance-scaler-eqn-0}
\end{equation}

\subsubsection{Inverting the Grad-Shafranov solution}

It is convenient to use Grad-Shafranov (GS) theory to obtain the zeroth order equilibrium.  This approach gives control of the axisymmetric plasma shape, and also benefits from existing understanding of the equation and its numerical solution.  A solution of the GS equation is the poloidal flux function $\Psi(R, z)$ is obtained from a given pressure function $p$, and the poloidal flux function $G$.  From these quantities, the corresponding profiles $I$ and $\iotaslash_0$, the current potential $K_0$ and coordinate mapping components $R_0$, $Z_0$ and $\nu_0$ can be calculated.  To perform the coordinate inversion, derivatives of the GS solution are computed, so a high degree of accuracy is needed.  A method is described in appendix \ref{grad-shafranov-appx}.

\subsection{${\cal O}(\epsilon^1)$}\label{ep1-sec}

As in \citet{plunk-helander-qas}, we do not modify the toroidal angle beyond zeroth order in the expansion ($\nu_1 = 0$, \etc, in Eqn.~\ref{phi-eqn}), but instead consider the corrections to the coordinate mapping to have a component in the $\phih$ direction, \ie

\begin{equation}
{\bf x}_1 = \Rh R_1(\theta, \psi, \varphi) + \zh Z_1(\theta, \psi, \varphi) + \phih \Phi_1(\theta, \psi, \varphi),\label{x1-def}
\end{equation}
from which it follows that 

\begin{eqnarray}
&\frac{\partial {\bf x}_1}{\partial \rho} &= \Rh \left(\frac{\partial R_1}{\partial \rho} + \Phi_1 \frac{\partial \nu_0}{\partial \rho}\right) + \zh \frac{\partial Z_1}{\partial \rho} + \phih \left(\frac{\partial \Phi_1}{\partial \rho} - R_1 \frac{\partial \nu_0}{\partial \rho}\right) \label{de1-eq}\\
&\frac{\partial {\bf x}_1}{\partial \theta} &=  \Rh \left(\frac{\partial R_1}{\partial \theta} + \Phi_1 \frac{\partial \nu_0}{\partial \theta}\right) + \zh \frac{\partial Z_1}{\partial \theta} + \phih \left(\frac{\partial \Phi_1}{\partial \theta} - R_1 \frac{\partial \nu_0}{\partial \theta}\right) \label{de2-eq}\\
&\frac{\partial {\bf x}_1}{\partial \varphi} &= \Rh\left( \frac{\partial R_1}{\partial \varphi} - \Phi_1 \right) + \zh \frac{\partial Z_1}{\partial \varphi} + \phih \left(R_1 + \frac{\partial \Phi_1}{\partial \varphi} \right). \label{de3-eq}
\end{eqnarray}
As $\varphi$ is an ignorable coordinate in the properly formulated first order equilibrium equations, we will assume

\begin{eqnarray}
R_1 = \hat{R}_1(\theta, \psi) \exp(i N \varphi) + \mathrm{c.c.},\label{x1-N-def-1}\\
Z_1 = \hat{Z}_1(\theta, \psi) \exp(i N \varphi) + \mathrm{c.c.},\label{x1-N-def-2}\\
\Phi_1 = \hat{\Phi}_1(\theta, \psi) \exp(i N \varphi) + \mathrm{c.c.},\label{x1-N-def-3}\\
\Kb_1 = \hat{\Kb}_1(\theta, \psi) \exp(i N \varphi) + \mathrm{c.c.}\label{x1-N-def-4},
\end{eqnarray}
with $N \neq 0$ an integer.  The deformation is thus non-axisymmetric, and the axisymmetric ($\varphi$-averaged) part of the local MHD constraint \ref{local-MHD-eqn} is $\iotaslash_1( \Gb g_{22}^{(0)} - \Ib g_{23}^{(0)} ) = 0$, from which we conclude that

\begin{equation}
\iotaslash_1 = 0.
\end{equation}

The first order MHD constraint is then

\begin{align}
\rho\left(\frac{\partial {\bf x}_1}{\partial \varphi} + \iotaslash_0 \frac{\partial {\bf x}_1}{\partial \theta}\right) =  & \;\Gb \left(\frac{\partial {\bf x}_0}{\partial \rho}\times\frac{\partial {\bf x}_1}{\partial \theta} +   \frac{\partial {\bf x}_1}{\partial \rho}\times\frac{\partial {\bf x}_0}{\partial \theta}\right) + \Ib \left(\frac{\partial {\bf x}_0}{\partial \varphi}\times\frac{\partial {\bf x}_1}{\partial \rho} + \frac{\partial {\bf x}_1}{\partial \varphi}\times\frac{\partial {\bf x}_0}{\partial \rho}\right) \nonumber\\ &+ \Kb_0 \left(\frac{\partial {\bf x}_0}{\partial \theta}\times\frac{\partial {\bf x}_1}{\partial \varphi} + \frac{\partial {\bf x}_1}{\partial \theta}\times\frac{\partial {\bf x}_0}{\partial \varphi}\right) + \Kb_1 \frac{\partial {\bf x}_0}{\partial \theta}\times\frac{\partial {\bf x}_0}{\partial \varphi}.\label{the-MHD-1-eqn}
\end{align}
What is needed is the $\exp(i N \varphi)$ component of this equation, obtained by substituting Eqns.~\ref{x1-N-def-1}-\ref{x1-N-def-4} into ${\bf x}_1$, Eqn.~\ref{x1-def}, and its derivatives, Eqns.~\ref{de1-eq}-\ref{de3-eq}.  The further substitution of zeroth order expressions, Eqns.~\ref{e10-eq}-\ref{e30-eq}, and projection along the cylindrical unit vectors, then yields a set of three equations for the unknowns $\hat{R}_1$, $\hat{Z}_1$, $\hat{\Phi}_1$, $\hat{\Kb}_1$ in terms of the known zeroth order solutions $R_0$, $Z_0$, $\nu_0$ and $\Kb_0$.  The system is completed with the force balance equation,

\begin{equation}
- i N \hat{\Kb}_1 - \iotaslash_0\frac{\partial \hat{\Kb}_1}{\partial \theta} + \hat{\Jb}_1 \mu_0 \frac{d\pb}{d\rho} = 0\label{f-balance-scaler-eqn-1}
\end{equation}
The $\exp(i N \varphi)$ component of the first order Jacobian, $\hat{\Jb}_1$, is obtained from Eqn.~\ref{Jb-eqn} by substituting Eqns~\ref{e20-eq}-\ref{e30-eq} and Eqns~\ref{de2-eq}-\ref{de3-eq}, into the following expression:

\begin{align}
\Jb_1 = 2\frac{\left(\frac{\partial {\bf x}_0}{\partial \varphi} + \iotaslash_0 \frac{\partial {\bf x}_0}{\partial \theta}\right)\cdot\left(\frac{\partial {\bf x}_1}{\partial \varphi} + \iotaslash_0 \frac{\partial {\bf x}_1}{\partial \theta}\right)}{\Gb + \iotaslash_0 \Ib}.
\end{align}

We note that QAS implies that $\hat{\Jb}_1 = 0$, so force balance on any QAS surface reduces to $i N \hat{\Kb}_1 + \iotaslash_0 \partial \hat{\Kb}_1/\partial \theta = 0$, which, assuming irrational $\iotaslash_0$, implies 

\begin{equation}
\hat{\Kb}_1 = 0.\label{QAS-fb-cond}
\end{equation}
On magnetic surfaces where QAS is not satisfied, the possibility of resonances in Eqn.~\ref{f-balance-scaler-eqn-1} must be considered.  It is easy to see that the equation can be uniquely solved for $\hat{\Kb}_1$, periodic in $\theta$, if $\iotaslash_0$ is not equal to a rational number.  Actually, some rational numbers are resonant, and some are not, in particular there are resonances at any magnetic surface where $\iotaslash_0$ satisfies

\begin{equation}
\iotaslash_0 = \frac{N}{m},
\end{equation}
for arbitrary integer $m$.  One strategy to avoid resonances is to constrain $\iotaslash_0$ to lie between two neighboring singular values.  In that case, force balance can be considered ``soluble'' throughout the plasma volume.  Note that, assuming $\iotaslash_0 \sim 1$, such ``safe'' ranges becomes increasingly narrow at large $N$, although resonances may be considered "high order" in this limit, and therefore less likely to pollute the solution..

Note that Eqn.~\ref{QAS-fb-cond} demonstrates that force balance is non-resonant on a QAS magnetic surface.  Actually, this reflects a general non-perturbative property, which follows directly from exact force balance and the relationship between $B$ and $\Jb$ (Eqns.~\ref{f-balance-scaler-eqn} and \ref{JacB-eqn-1}), namely that quasi-symmetry drastically simplifies the source term in force balance (the Fourier series of $\Jb$ in $\theta$ and $\varphi$ has only terms of a single helicity), so that an MHD equilibrium that is quasi-symmetric globally (on all magnetic surfaces) should be free from nontrivial resonances; see also \citet{burby2019mathematics} and \citet{Rodr_guez_2020}.

To summarize, at first order the equations to be solved are the three components of Eqn.~\ref{the-MHD-1-eqn}, coupled with force balance, Eqn.~\ref{f-balance-scaler-eqn-1}, for the four unknown functions $\hat{R}_1(\rho, \theta)$, $\hat{Z}_1(\rho, \theta)$, $\hat{\Phi}_1(\rho, \theta)$, $\hat{\Kb}_1(\rho, \theta)$.  The domain is the unit disk, $\rho \in [0, 1]$ and the boundary condition is QAS, which translates to $\hat{\Kb}(\rho, \theta)|_{\rho = 1} = 0$.  No rotational transform is obtained at this order, but the first order solution does generally induce transform at ${\cal O}(\epsilon^2)$, \ie enters the computation of $\iotaslash_2$.

\subsection{${\cal O}(\epsilon^2)$}
Proceeding to the next order, the equations are quite similar as before, but now include terms that are quadratic in first order quantities.  The second order coordinate mapping is

\begin{equation}
{\bf x}_2 = \Rh R_2(\theta, \psi, \varphi) + \zh Z_2(\theta, \psi, \varphi) + \phih \Phi_2(\theta, \psi, \varphi),\label{x2-def}
\end{equation}
and its derivatives are

\begin{eqnarray}
&\frac{\partial {\bf x}_2}{\partial \rho} &= \Rh \left(\frac{\partial R_2}{\partial \rho} + \Phi_2 \frac{\partial \nu_0}{\partial \rho}\right) + \zh \frac{\partial Z_2}{\partial \rho} + \phih \left(\frac{\partial \Phi_2}{\partial \rho} - R_2 \frac{\partial \nu_0}{\partial \rho}\right) \label{de1-eq2}\\
&\frac{\partial {\bf x}_2}{\partial \theta} &=  \Rh \left(\frac{\partial R_2}{\partial \theta} + \Phi_2 \frac{\partial \nu_0}{\partial \theta}\right) + \zh \frac{\partial Z_2}{\partial \theta} + \phih \left(\frac{\partial \Phi_2}{\partial \theta} - R_2 \frac{\partial \nu_0}{\partial \theta}\right) \label{de2-eq2}\\
&\frac{\partial {\bf x}_2}{\partial \varphi} &= \Rh\left( \frac{\partial R_2}{\partial \varphi} - \Phi_2 \right) + \zh \frac{\partial Z_2}{\partial \varphi} + \phih \left(R_2 + \frac{\partial \Phi_2}{\partial \varphi} \right). \label{de3-eq2}
\end{eqnarray}
The appearance of the nonlinear terms occurs (in the MHD constraint and force balance equation) at toroidal mode numbers $\pm 2N$, and also in the axisymmetric component, which now must be solved to obtain $\iotaslash_2$.  We note that the appearance of toroidal mode numbers $\pm 2N$ at second order implies a denser set of possible force balance resonances at higher orders in the expansion, \ie $\iotaslash_0 = 2N/m$, which may justify further restriction on the chosen profile for $\iotaslash_0$.  Even if the problem will only be solved at first order, higher order resonances may occur in the exact force balance equation that must be satisfied by the full equilibrium.

In the vacuum case \citep{plunk-helander-qas}, $\iotaslash_2$ was obtained as a solubility constraint of the axisymmetric component ($\varphi$-average) of the local MHD constraint, Eqn.~\ref{local-MHD-eqn}.  This result does not appear to generalize in a simple way, implying that the full system (MHD constraint plus force balance) must be solved to obtain $\iotaslash_2$.

\section{Numerical Solution}\label{numerical-sec}

The task now is to solve the system composed of Eqn.~\ref{the-MHD-1-eqn} and Eqn.~\ref{f-balance-scaler-eqn-1} for the unknowns $\hat{R}_1$, $\hat{Z}_1$, $\hat{\Phi}_1$, $\hat{\Kb}_1$, subject to QAS ($\hat{\Kb} = 0$) on a specified magnetic surface, typically the outermost magnetic surface ($\rho = 1$).  Note that this boundary surface need not necessarily be taken to be located at the plasma edge.  It has been suggested \citep{Henneberg_2019} that it may be optimal to satisfy QAS at some intermediate magnetic surface, which can be implemented here by redefinition of the coordinate $\rho$, or simply choosing a boundary $\rho < 1$.  Henceforth we assume $\rho = 1$ for simplicity.

The ``pseudo-cartesian'' coordinates $\xpc = \rho \cos(\theta)$ and $\ypc = \rho \sin(\theta)$ are used for numerical purposes, instead of the polar coordinates $\rho$ and $\theta$.  These have the advantage that they do not possess the singularity of the polar coordinates ($\rho$, $\theta$) as $\rho \rightarrow 0$, and they do not require periodicity to be enforced in $\theta$, or any analyticity at $\rho = 0$.  The only advantage found in using $\rho$-$\theta$ coordinates is to explicitly observe development of non-analyticity in the solutions on resonant surfaces (satisfying $\iotaslash_0 = N/m$).

The finite element method is used, reformulating the problem as an equivalent ``least squares'' problem.  The least squares finite element method offers better convergence and stability properties for systems of first order PDEs \citep{Jiang-FEM-book}.  To show how the problem is reformulated, we introduce the vector field ${\bf u}(\xpc, \ypc) = [\hat{\Kb}_1, \hat{R}_1, \hat{Z}_1, \hat{\Phi}_1]^T$, together with the inner product

\begin{equation}
\langle {\bf v} | {\bf u} \rangle = \int d\xpc d\ypc {\bf v}^*\cdot{\bf u},
\end{equation}
where ${\bf v}^*$ denotes the complex conjugate of ${\bf v}$, and the integral is performed over the computational domain, the unit disk $\Omega$.  The original first order system of equations (\ie the $\Rh$, $\phih$, and $\zh$ components of Eqn.~\ref{the-MHD-1-eqn}, coupled with Eqn.~\ref{f-balance-scaler-eqn-1}) can be written as ${\cal L} {\bf u} = 0$, where

\begin{equation}
[{\cal L} {\bf u}]_i = a_{ij} u_j + \alpha_{ijk} \partial_j u_k,
\end{equation}
where $u_j$ denotes the $j$th component of ${\bf u}$, $\partial_1$ and $\partial_2$ denote $\partial/\partial \xpc$ and $\partial/\partial \ypc$, respectively, and the tensors $a_{ij}$ and $\alpha_{ijk}$ encode the coefficients of the system of equations.  The adjoint of ${\cal L}$ is denoted as ${\cal L}^\dagger$, and is given by

\begin{equation}
[{\cal L}^\dagger {\bf u}]_i = a_{ji}^* u_j - \partial_j(\alpha_{kji}^* u_k).
\end{equation}
With these definitions, our problem is transformed into solving the following eigenvalue problem

\begin{equation}
{\cal L}^\dagger {\cal L} {\bf u} = \lambda {\bf u},
\end{equation}
subject to the QAS boundary condition $u_1 = 0$ on the boundary $\partial\Omega$, for solutions with eigenvalue $\lambda \rightarrow 0$.  This system is generated by computer algebra, and not explicitly written down, due to its complexity.

\subsection{Examples}

\begin{figure}
    \centering
    \begin{subfigure}[t]{0.55\textwidth}
        \centering
\includegraphics[width=\textwidth]{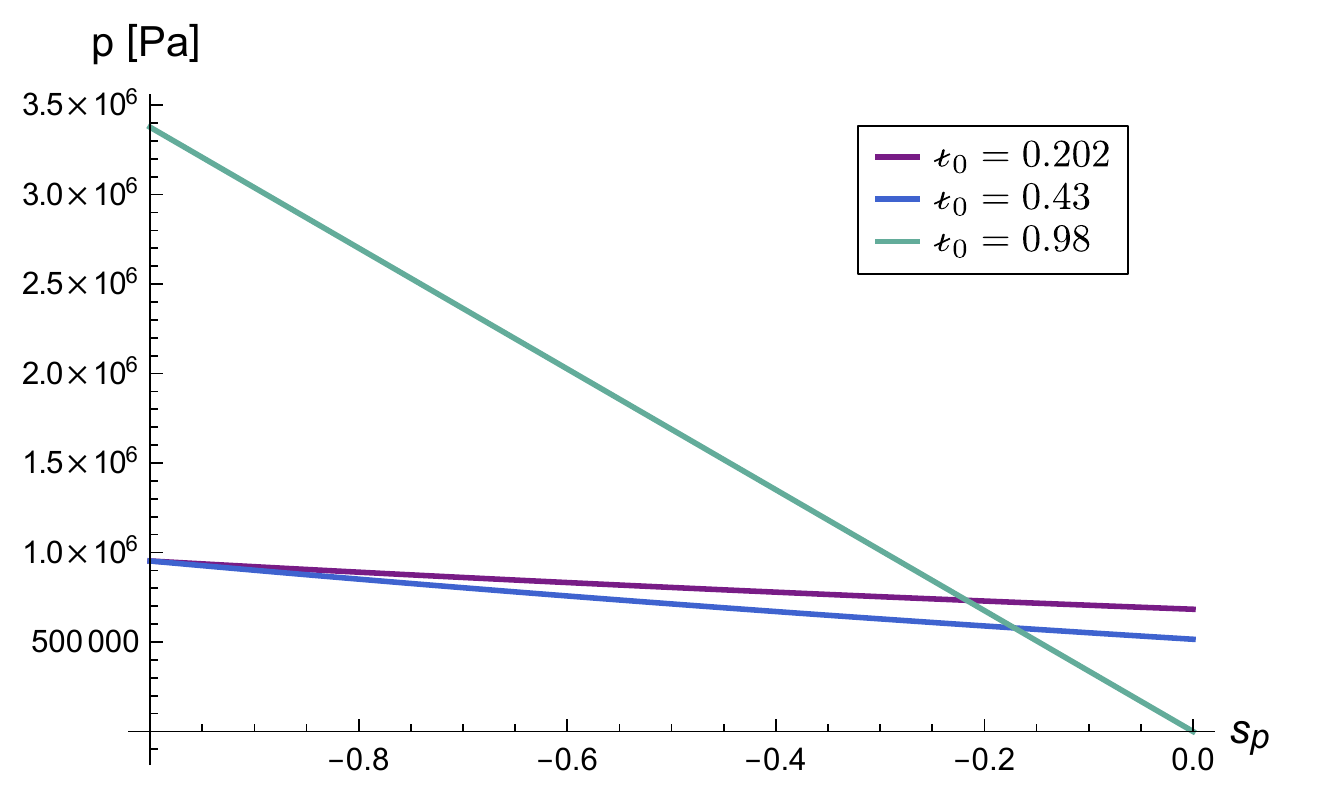}
        \caption{Pressure profile}
    \end{subfigure}%
    ~ 
    \begin{subfigure}[t]{0.45\textwidth}
        \centering
\includegraphics[width=\textwidth]{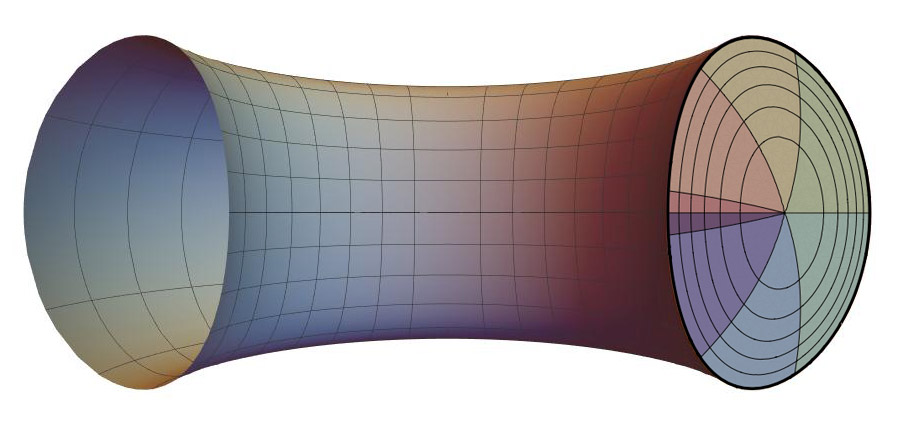}
        \caption{Outer magnetic surface shape}
    \end{subfigure}
\caption{ITER-like equilibria with constant rotational transform profiles.  The pressure profile (a) is plotted versus normalized poloidal flux $s_p = \Psi/|\Psi_{\mathrm axis}|$ where $\Psi_{\mathrm axis}$ is the value on axis, and $\Psi$ is taken to be zero at the outermost surface.  Note that the mesh on the magnetic surface is made of lines of constant geometric angle $\phi$ and constant Boozer angle $\theta$; the end cap shows lines of constant $\theta$ and $\Psi$.}
\label{iter-like-equilib-fig}
\end{figure}

To demonstrate that the above numerical method works, in practice, and give a flavor of possible solutions, we consider perturbations of two tokamak equilibria, based on ITER.  The ITER-like equilibria have their outer surface shape defined by the Solev'ev equilibrium given in \citet{pataki-cerfon}, but scaled up so that the magnetic axis has a radial position of $6.68$ m and the total toroidal flux over $2\pi$ of $15.7$ Weber.  The model pressure profiles that are linear in the poloidal flux function $\Psi$, as shown in figure \ref{iter-like-equilib-fig}, and three different constant rotational transform profiles are considered, $\iotaslash_0 = 0.202$, $0.47$ and  $0.98$.  These values are chosen to avoid resonances for $N = 2$, $4$ and $8$; see section \ref{ep1-sec}.

\begin{figure}
    \centering
    \begin{subfigure}[t]{0.3\textwidth}
        \centering
\includegraphics[width=\textwidth]{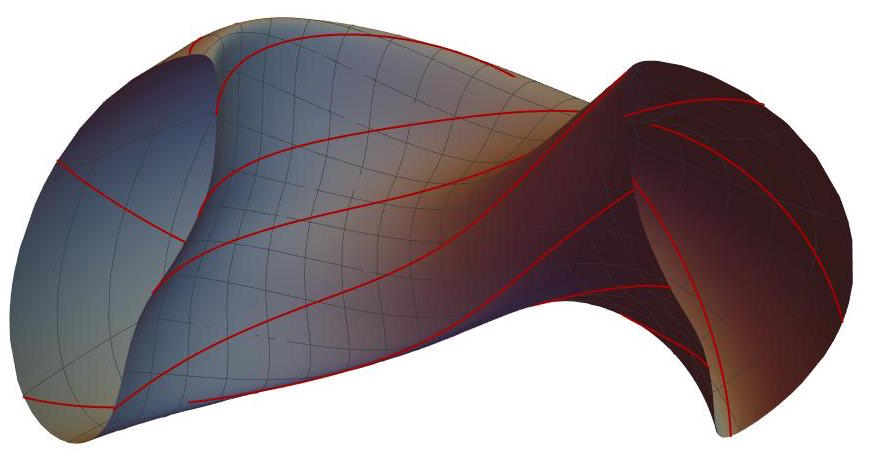}
    \end{subfigure}
    ~ 
    \begin{subfigure}[t]{0.3\textwidth}
        \centering
\includegraphics[width=\textwidth]{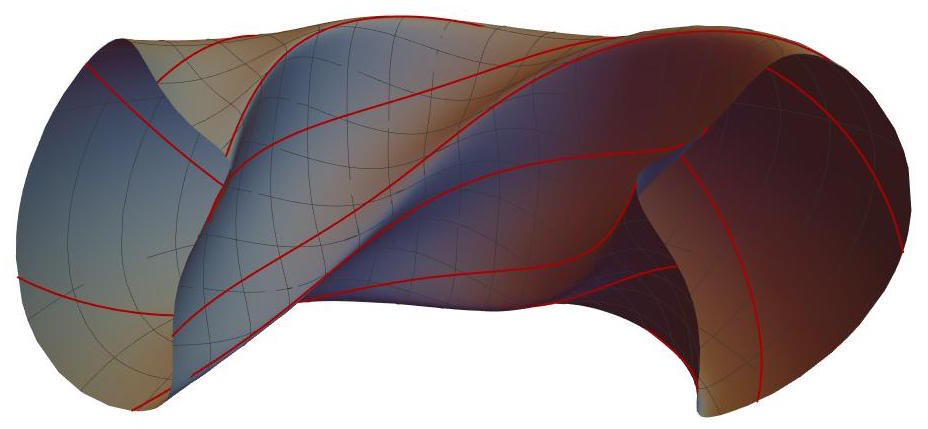}
    \end{subfigure}
    ~ 
    \begin{subfigure}[t]{0.3\textwidth}
        \centering
\includegraphics[width=\textwidth]{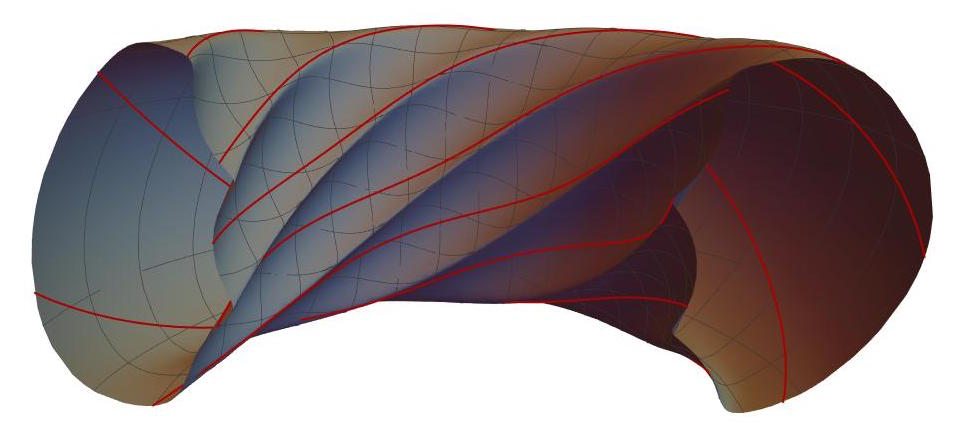}
    \end{subfigure}
    \\
    \begin{subfigure}[t]{0.3\textwidth}
        \centering
\includegraphics[width=\textwidth]{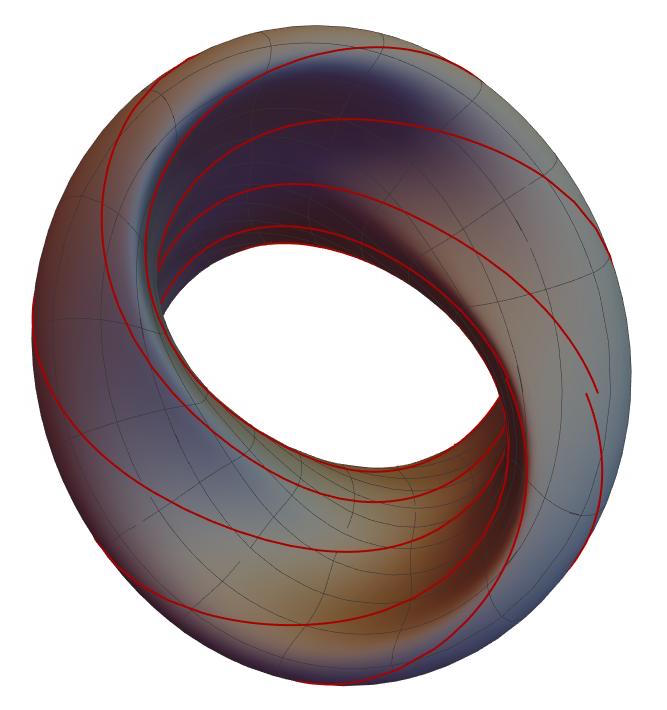}
        \caption{$N = 2$}
\label{iota-98-fig-a}
    \end{subfigure}
    ~ 
    \begin{subfigure}[t]{0.3\textwidth}
        \centering
\includegraphics[width=\textwidth]{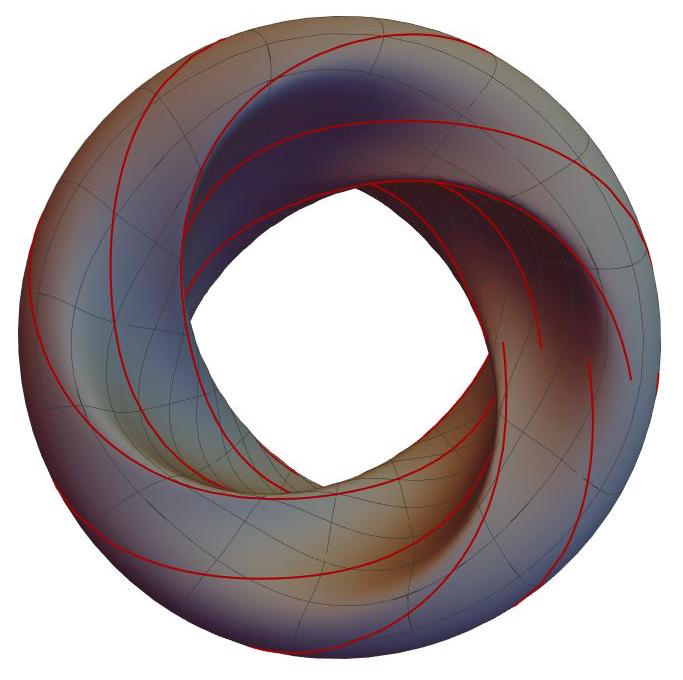}
        \caption{$N = 4$}
\label{iota-98-fig-b}
    \end{subfigure}
    ~ 
    \begin{subfigure}[t]{0.3\textwidth}
        \centering
\includegraphics[width=\textwidth]{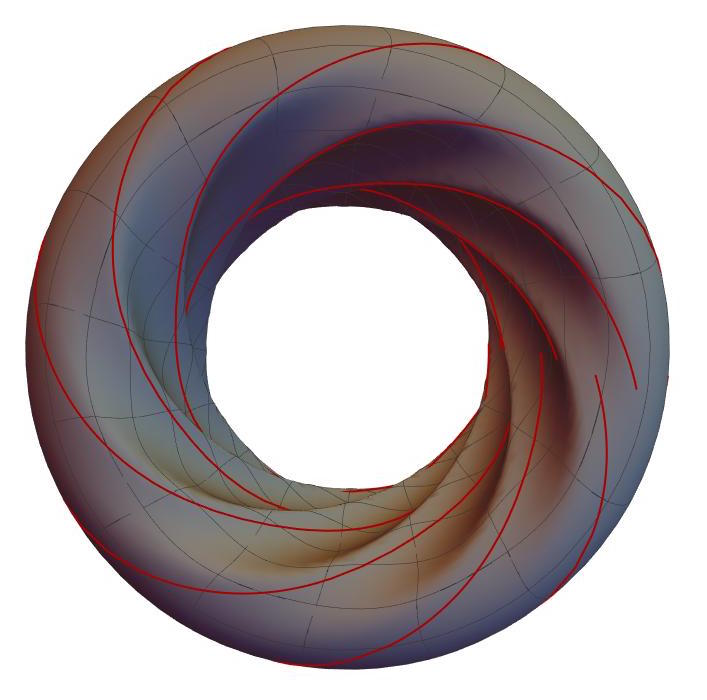}
        \caption{$N = 8$}
\label{iota-98-fig-c}
    \end{subfigure}
\caption{Outer magnetic surface shapes for ITER-like QAS equilibria with near-unity rotational transform, $\iotaslash_0 = 0.98$.  Two views angles are shown, from the top and from the side, with the side view showing half a toroidal turn; a sample of field line segments are plotted in red.  The mesh on the magnetic surface correspond to lines of constant Boozer angles, $\theta$ and $\varphi$.}
\label{iota-98-fig}
\end{figure}

To independently evaluate the first order QAS numerical solutions, the outer surface shape can be generated and provided to the VMEC code \citep{Hirshman} as input for a fully nonlinear calculation, as was done in \citet{plunk-helander-qas}, and the result then passed to the BOOZ\_XFORM code \citep{booz-xform} to check the level of QAS as predicted by the theory.  To produce the surface shape, the perturbation amplitude is controlled via the arbitrary small parameter $\epsilon$ in ${\bf x} \approx {\bf x}_0 + \epsilon {\bf x}_1$.  Three such surfaces are shown in Fig.~\ref{iota-98-fig}.

\begin{figure}
\includegraphics[width=0.55\textwidth]{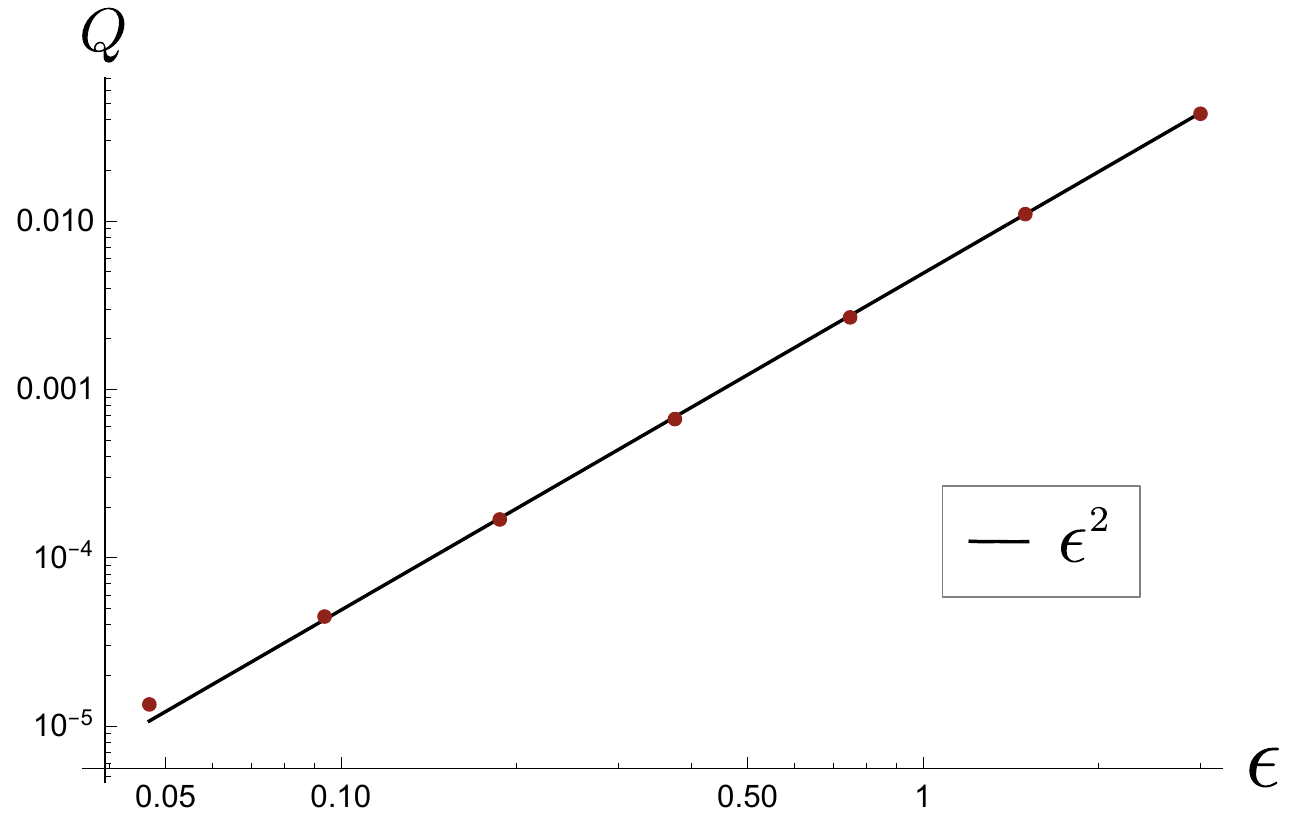}
\includegraphics[width=0.45\textwidth]{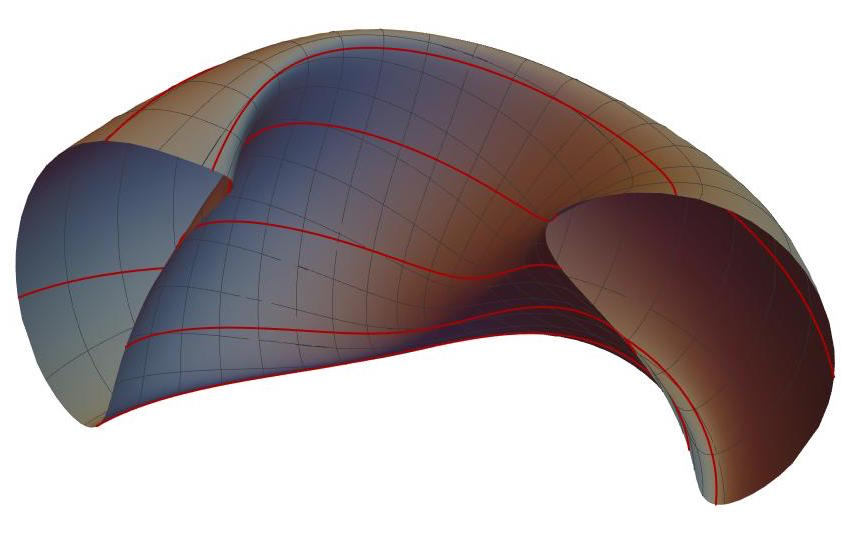}
\caption{Example of perturbed tokamak equilibrium ($N = 2$) with ITER-like shaping.  Unperturbed rotational transform is $\iotaslash_0 = 0.202$ at all radial locations.  Left: Theoretical scaling of $\epsilon^2$ is well satisfied for QAS error, defined in Eqn.~\ref{Q-eqn}.  Right: Outer surface shape visualized for case of strongest shaping (largest perturbation), with $1$ toroidal field period plotted, and a sample of field line segments.}
\label{ITER-202-example-fig}
\end{figure}

The solutions are valid to first order in the expansion, and it can therefore be expected that the error in QAS should scale as $\epsilon^2$, the confirmation of which is shown for one case in Fig.~\ref{ITER-202-example-fig}, where the error is measured as follows:

\begin{equation}
Q = \frac{\left(\sum_{m, n\neq 0}|\hat{B}_{mn}|^2\right)^{1/2}}{\left(\sum_{m, n}|\hat{B}_{mn}|^2\right)^{1/2}},\label{Q-eqn}
\end{equation}
where $\hat{B}_{mn}$ is the Fourier coefficient of $|{\bf B}|$ calculated in the Boozer angles by the BOOZ\_XFORM.  It should be noted that not all of the cases reported here match so closely with the theoretical scaling.  Some have only a limited range at larger values of $\epsilon$ where the quadratic scaling is observed, and exhibiting the weaker $\epsilon^1$ scaling for smaller values of $\epsilon$, associated with non-QAS perturbations.  Although it is expected, for instance, that numerical error in the first order solution can introduce $\epsilon^1$ scaling which must dominate at sufficiently small $\epsilon$, it does not appear that the linear scaling observed here is related to numerical error in the three codes being used, of the type introduced by finite resolution.  It is therefore suspected that a more fundamental issue is at fault, for instance (1) the presence of force balance singularities in the fully nonlinear calculation performed by VMEC (which are formally absent from our first order calculation of ${\bf x}_1$), or (2) the possibility that the problem we are solving (QAS on a single surface of non-axisymmetric perturbations) is sometimes (or always) not well posed; this issue will be investigated in future work.  Nevertheless, the low observed QAS error in solutions obtained so far indicate that the numerical method developed here should be practically useful.

\begin{figure}
    \centering
    \begin{subfigure}[t]{0.3\textwidth}
        \centering
\includegraphics[width=\textwidth]{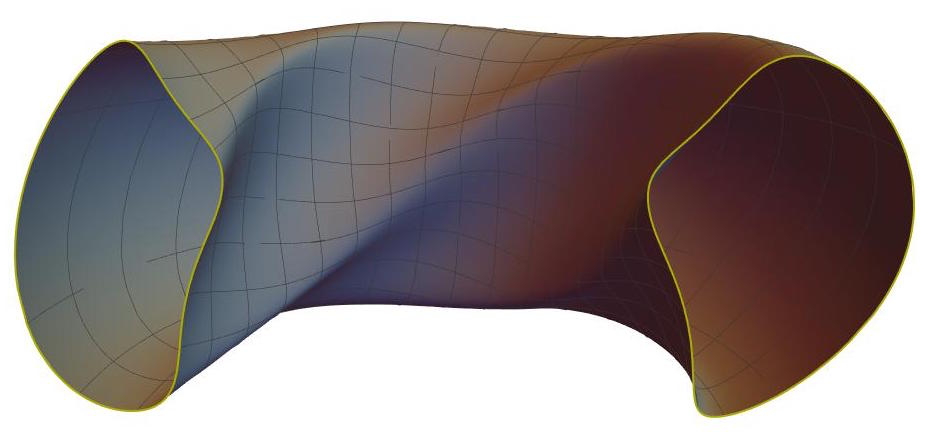}
        \caption{$\epsilon = 0.5$}
    \end{subfigure}%
    ~ 
    \begin{subfigure}[t]{0.3\textwidth}
        \centering
\includegraphics[width=\textwidth]{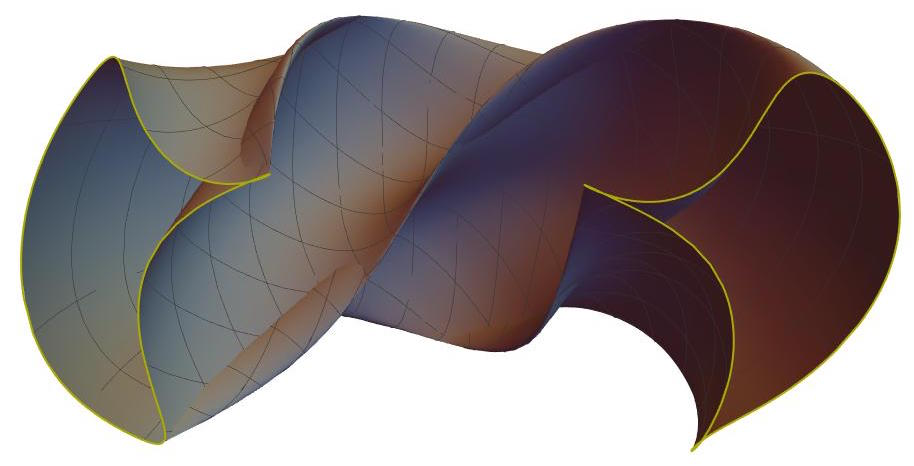}
        \caption{$\epsilon = 1.8$}
    \end{subfigure}
    ~ 
    \begin{subfigure}[t]{0.3\textwidth}
        \centering
\includegraphics[width=\textwidth]{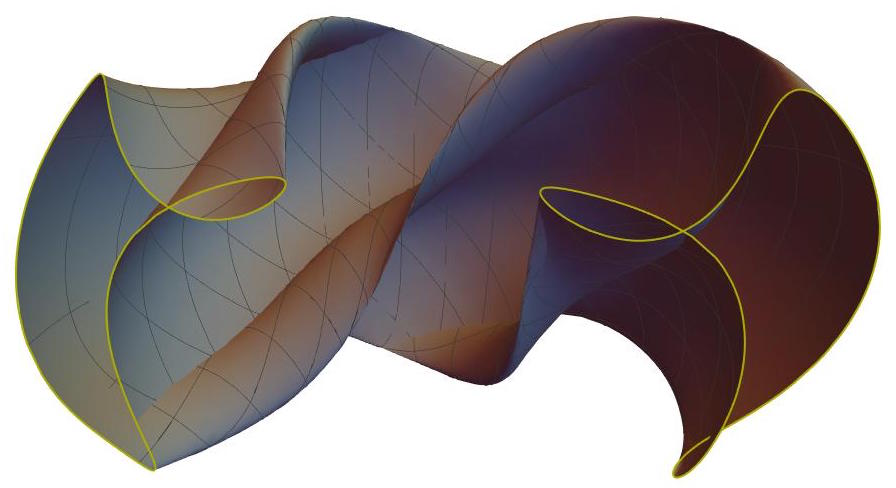}
        \caption{$\epsilon = 2.5$}
    \end{subfigure}
\caption{As the perturbation amplitude $\epsilon$ is increased, the magnetic surfaces ``reconnect'', invalidating the solution.  Here is an example, with $\iotaslash_0 = 0.98$ and $N=4$, showing a single field period.  For comparison, the corresponding case in table \ref{iter-table}, and plotted in figure \ref{iota-98-fig-b} corresponds to $\epsilon = 0.75$.}
\label{surface-breaking-fig}
\end{figure}

One issue encountered with using the inverse representation of a magnetic equilibrium is that the coordinate mapping is not generally invertible for the magnetic coordinates.  Invertibility breaks down, when distinct points in the magnetic coordinate space, say ($\rho_1$, $\theta_1$, $\varphi_1$) and ($\rho_2$, $\theta_2$, $\varphi_2$), yield the same point in physical space, \eg ${\bf x}(\rho_1, \theta_1, \varphi_1) = {\bf x}(\rho_2, \theta_2, \varphi_2)$.  The QAS solutions here, being based on known axisymmetric equilibria will not suffer from this problem if the perturbation amplitude is chosen to be sufficiently small.  However, the problem can be reliably encountered at large values of $\epsilon$, as demonstrated in Fig.~\ref{surface-breaking-fig}.  What is remarkable about this phenomenon, which is associated with the perturbation overwhelming the zeroth order mapping, is that the theoretical QAS scaling tends to hold even as the singular point is approached, as demonstrated by Fig.~\ref{ITER-202-example-fig}.  Therefore, the VMEC solutions shown here are generally chosen to correspond to a value of $\epsilon$ close to the singular point, but not so large as to create sharp features in the outer magnetic surface that require more than $10-20$ Fourier harmonics to properly resolve in VMEC.  
 
Using the procedure described  above, the QAS solutions, though formally only perturbative, can yield strongly shaped plasma equilibria with reasonable level of QAS, and finite ``external'' rotational transform, as measured by the difference between the total rotational transform and that of the original axisymmetric equilibrium.  This is shown by table \ref{iter-table}, where a total of nine cases are described, corresponding to three toroidal mode numbers applied to the equilibria of three different values of constant rotational transform.  Each row of the table corresponds to a single value of $\epsilon$ (although a sequence of values were generally calculated to investigate scaling).  The fraction of external rotational transform generated by the perturbation is given in the third column:

\begin{equation}
f^\mathrm{ext}_{\iotaslash} = \frac{\iotaslash - \iotaslash_0}{\iotaslash}\bigg\rvert_{\rho = 1}
\end{equation}
Next is the root-mean-squared value of the modulus of the perturbed magnetic field, divided by zeroth order field strength, with the average performed over $\theta$ and $\varphi$, denoted $E$:

\begin{equation}
E = \left\langle\frac{|\delta {\bf B}|}{|{\bf B}_0|}\right\rangle_\mathrm{rms} = \left (\frac{\epsilon^2}{4\pi^2} \int d\varphi d\theta \frac{\left | \frac{\partial {\bf x}_1}{\partial \varphi} + \iotaslash_0 \frac{\partial {\bf x}_1}{\partial\theta} \right |^2}{\left | \frac{\partial {\bf x}_0}{\partial \varphi} + \iotaslash_0 \frac{\partial {\bf x}_0}{\partial\theta} \right |^2} \right)^{1/2},
\end{equation}
where we note that the above expression assumes the first order Jacobian to be zero (QAS).  This measure gives some sense of how strong the perturbation is, and may be used to estimate the size of external current distributions needed to achieve the total field.  The next column provides the QAS error, $Q$, defined in Eqn.~\ref{Q-eqn}.  The chosen values of $\epsilon$ are somewhat arbitrary, so it is useful to calculate normalized values to compare the various solutions.  For that reason we also give inferred values of the magnetic perturbation measures $E_{10}$ and $E_{15}$, that would be obtained for external rotational transforms of $10\%$ and $15\%$, respectively.  Analogous quantities for QAS error are denoted $Q_{10}$ and $Q_{15}$.  These are calculated for each case by using the fact that $\delta \iotaslash - \iotaslash_0$ scales theoretically as $\epsilon^2$ (confirmed for all cases), as does $Q$, whereas $E$ scales as $\epsilon^1$.  We stress that these values, being obtained from first order solutions, and not benefitting from any further optimization, should only be taken as a indicator of what can be achieved by perturbing an axisymmetric equilibrium.  However, what seems clear is that, despite the fact that the perturbation of the field $E$ scales as $\epsilon^1$ whereas external transform scales as $\epsilon^2$, significant values of the latter can still be achieved at modest values of the former, as for instance shown by the $\iotaslash_0 = 0.43$, $N=4$ case where the rms field strength fraction is not much larger that the external rotational transform fraction.

\begin{table}
\begin{center}
\begin{tabular}{ c  c  c  c  c | c c | c }
$\iotaslash_0$ & $N$ & $ f^\mathrm{ext}_{\iotaslash}$ & $E$ & $Q$ & $E_{10}/E_{15}$ & $Q_{10}/Q_{15}$ & Loc \\
\hline
$0.202$ & $2$ & $0.21$ &    $0.37$   & $0.043$ & $0.26/0.31$ &                 $0.021/0.031$ & \num{8.7e-2} \\
$0.202$ & $4$ & $0.34$ &    $0.30$   & $0.025$ & $0.16/0.20$ &                 $0.0074/0.011$ & \num{4.7e-3} \\
$0.202$ & $8$ & $0.16$ &    $0.32$   & $0.038$ & $0.25/0.31$ &                 $0.024/0.036$ & \num{3.4e-4} \\

$0.43$ &   $2$ & $0.29$ &  $0.31$   & $0.0087$ & $0.19/0.23$ &                 $0.0030/0.0045$  & \num{1.1e-1}\\
$0.43$ &   $4$ & $0.21$ &  $0.17$   & $0.0029$ & $0.12/0.14$ &                 $0.0014/0.0021$ & \num{1.3e-3}\\
$0.43$ &   $8$ & $0.13$ &  $0.14$   & $0.0034$ & $0.12/0.15$ &                 $0.0026/0.0039$ & \num{2.7e-4} \\

$0.98$ &   $2$ & $0.039$ &  $0.20$   & $0.019$ & $0.32/0.39$ &                 $0.050/0.074$ & \num{2.5e-1} \\
$0.98$ &   $4$ & $0.088$ &  $0.16$   & $0.0047$ & $0.17/0.21$ &               $0.0052/0.0079$ & \num{4.2e-2} \\
$0.98$ &   $8$ & $0.078$ &  $0.11$   & $0.0046$ & $0.13/0.15$ &               $0.0059/0.0088$ & \num{3.7e-3} \\
\end{tabular}
\caption{Summary of results for ITER-like QAS equilibria}
\label{iter-table}
\end{center}
\end{table}

An interesting qualitative feature of the QAS perturbations is the tendency to ``localize'' to the inboard side (\eg at lower values of the radial coordinate $R$), in the sense that the amplitude of the perturbed magnetic field is larger there than on the outboard.  This is quantified in the final column of the table \ref{iter-table}, labeled ``Loc'', where we calculate the ratio of the root-mean-squared value of the perturbed magnetic field ${\bf \delta B}$ on the outboard (defined such that $\theta = 0$) to inboard ($\theta = \pi$), where the average is done only over the toroidal angle $\varphi$.  This feature is more pronounced at larger $N$ and lower aspect ratio, as was also observed for the vacuum case (see the appendix of \citet{plunk-helander-qas}; the explanation here may be similar).  We note that the high-$N$ perturbations also only weakly penetrates radially into the plasma, as the rotational transform $\iotaslash$ can be observed to fall rapidly, from the edge, to the unperturbed value $\iotaslash_0$. 

\section{Conclusion}

This work gives the first set of results showing the direct construction of QAS perturbations of non-trivial axisymmetric plasma equilibria.  It has been demonstrated that, despite the perturbative nature of the calculation, relatively strongly shaped stellarator equilibria may be obtained with significant external rotational transform ($10-15\%$), at a similar level of average perturbed magnetic field.  This finding agrees with a previous study \citep{ku-boozer} using conventional numerical optimization.  However, the method of the present work allows for a more extensive exploration of the design space, as the general QAS perturbation corresponds to a sum of modes with suitably non-resonant toroidal mode numbers.  This potentially opens new avenues for exploring the concept of a stellarator-tokamak hybrid device.  

One interesting initial finding is that relatively high-$N$ perturbations (here as high as $N=8$) seem to efficiently produce finite external rotational transform (\eg $10\%$), while diminishing strongly in amplitude both radially and polloidally, and showing very good satisfaction of QAS, much less than $1\%$ error.  Such perturbations may be generated by a more ``modest'' distribution of coils localized to the inboard side of the plasma.  Additionally, with the perturbation localized to the high-field side of the device, it should only significantly affect the radial drift of barely trapped particles, rendering the overall neoclassical transport especially small.

One benefit of perturbative studies like the present is the ability to characterize the size of the solution space of optimal stellarators.  Similar to what was found by \citet{plunk-helander-qas}, we conclude that the freedom in QAS designs comes from (1) the zeroth order equilibrium, which is in this case includes plasma profiles in addition to the two-dimensional shaping (\eg triangularity, elongation, aspect ratio, \etc), and (2) the solution space of the QAS perturbation.  For the latter, there are also some differences: first, it appears that, for fixed toroidal mode number $N$, the solution is unique, whereas \citet{plunk-helander-qas} found that solutions came in pairs -- the latter situation may stem from the symmetry of the $\iotaslash_0 = 0$ scenario; we note that this limit cannot be approached within the current framework, as the resonant values of $\iotaslash$ become dense near $\iotaslash = 0$.  Second, the choice of toroidal mode number $N$ is constrained, at least formally, by the profile $\iotaslash_0(\rho)$, in the sense that resonances ($\iotaslash_0 = N/m$) must be avoided to guarantee smooth solutions at first order, with further resonances might be considered at higher order.  Therefore, the realizable solution space for QAS perturbations of a given axisymmetric equilibrium may be limited to a small number of toroidal mode numbers.  Such a small space might be rapidly explored to identify QAS equilibria that satisfy additional requirements.

The success demonstrated here in directly constructing QAS solutions with an inverse method, using Boozer coordinates, gives some hope that the fully nonlinear problem may be formulated and solved in a similar fashion, \ie with a code similar to VMEC that would obtain quasi-symmetric stellarator equilibria directly, and without approximations.  To accomplish this, it is necessary to identify the appropriate amount of boundary information to yield a well-posed problem; the findings of this paper should provide a useful guide in this endeavor.

{\bf Acknowledgements.}  Thanks to P. Helander and S. Henneberg for useful discussions, and to S. Lazerson, E. Strumberger and J. Geiger for help with VMEC.  Thanks also to Y. Turkin and P. Aleynikov for help with tokamak equilibrium files.

\bibliographystyle{unsrtnat}
\bibliography{Quasisymmetry-near-axisymmetry-2.bib}

\appendix
\section{Magnetic geometry}\label{geometry-appx}

From the coordinates ($\rho$, $\theta$, $\varphi$) we define local basis vectors ($\bnabla\rho$, $\bnabla\theta$, $\bnabla\varphi$) and ($\eb_1 = \partial {\bf x}/\partial \rho$, $\eb_2 = \partial {\bf x}/\partial \theta$, $\eb_3 = \partial {\bf x}/\partial \varphi$).  The metric components are defined in the usual way

\begin{equation}
g_{ij} \equiv \eb_i\cdot\eb_j,
\end{equation}

\noindent and the Jacobian for these coordinates is

\begin{equation}
\Jac = \frac{1}{\bnabla\rho\cdot(\bnabla\theta\times\bnabla\varphi)} = \eb_1\cdot (\eb_2\times\eb_3)
\end{equation}

\noindent Additionally, assigning $(u_1, u_2, u_3) = (\rho, \theta, \varphi)$, we see that $\eb_i\cdot\bnabla u_j = \delta_{ij}$, and the following identities are easily verified

\begin{eqnarray}
\eb_1 = \Jac(\bnabla u_2\times\bnabla u_3),\quad \eb_2 = \Jac(\bnabla u_3\times\bnabla u_1),\quad \eb_3 = \Jac(\bnabla u_1\times\bnabla u_2),\\
\bnabla u_1 = \frac{\eb_2\times\eb_3}{\Jac},\quad \bnabla u_2 = \frac{\eb_3\times\eb_1}{\Jac},\quad \bnabla u_3 = \frac{\eb_1\times\eb_2}{\Jac}.
\end{eqnarray}

\section{Useful forms of $\Jac$ and $B$}

Taking $B^2 = \B_{\mathrm{cov}} \cdot \B_{\mathrm{con}}$ gives

\begin{equation}
\Jb \bar{B}^2 = \Gb + \iotaslash \Ib,\label{JacB-eqn-1}
\end{equation}

\noindent where we recall the definition $\bar{B} = B/(2\psi_b)$.  Taking $B^2 = \B_{\mathrm{con}} \cdot \B_{\mathrm{con}}$ gives

\begin{equation}
\Jb^2 \bar{B}^2 = |\eb_3 + \iotaslash \eb_2|^2 = g_{33} + 2\iotaslash g_{23} + \iotaslash^2 g_{22}.\label{JacB-eqn-2}
\end{equation}

\noindent Using Eqn.~\ref{JacB-eqn-1}-\ref{JacB-eqn-2} we can express the magnetic field strength ``locally'' (in terms of only surface metrics)

\begin{equation}
\frac{(\Gb+\iotaslash \Ib)^2}{\bar{B}^2} = g_{33} + 2 \iotaslash g_{23} + \iotaslash^2 g_{22}.\label{B-local-eqn}
\end{equation}

\noindent Using Eqn.~\ref{JacB-eqn-1}-\ref{JacB-eqn-2} we can also express the Jacobian locally

\begin{equation}
\Jb (\Gb + \iotaslash \Ib) = g_{33} + 2 \iotaslash g_{23} + \iotaslash^2 g_{22}.\label{Jac-local-eqn}
\end{equation}

\section{The MHD constraint}

One can write the constraint Eqn.~\ref{the-MHD-eqn}, in different ways.  Taking the $\eb_1$, $\eb_2$, and $\eb_3$ components of this equation gives

\begin{eqnarray}
g_{13} + \iotaslash g_{12} = \Kb \Jb,\label{e1-eqn}\\
g_{23} + \iotaslash g_{22} = \Ib \Jb,\label{e2-eqn}\\
g_{33} + \iotaslash g_{23} = \Gb \Jb.\label{e3-eqn}
\end{eqnarray}

Note that Eqns.~\ref{e2-eqn} and \ref{e3-eqn} only involve surface metrics, and may be combined, eliminating the Jacobian, to obtain the local MHD constraint:

\begin{equation}
\Ib \left(g_{33} + \iotaslash g_{23}\right) = \Gb \left(g_{23} + \iotaslash g_{22}\right).\label{local-MHD-eqn}
\end{equation}
Combining the $\eb_i$ components we can derive three metric constraints not explicitly involving the Jacobian $\Jac$, the one above and the following two

\begin{eqnarray}
\Ib \left(g_{13} + \iotaslash g_{12}\right) = \Kb \left(g_{23} + \iotaslash g_{22}\right),\label{Jfree-MHD-eqn-2}\\
\Kb \left(g_{33} + \iotaslash g_{23}\right) = \Gb \left(g_{13} + \iotaslash g_{12}\right).\label{Jfree-MHD-eqn-3}
\end{eqnarray}

\noindent Note that this system is incomplete for a vacuum field because then $K = 0$ and $I = 0$ and Eqn.~\ref{Jfree-MHD-eqn-2} provides no information.  It can then be completed by including Eqn.~\ref{e3-eqn}.

\section{Zeroth order axisymmetric equilibrium by direct formulation}\label{grad-shafranov-appx}

Here we consider magnetic coordinates $\psi$, $\theta$, $\varphi$ as functions of cylindrical coordinates $R$, $Z$ and $\phi$.  The condition of axi-symmetry can be stated as the condition that the $\hat{\bf R}(\phi)$, $\hat{\bphi}(\phi)$ and $\hat{\bf z}$ components of $\B$ are independent of $\phi$.  This implies that the magnetic surfaces must be themselves axisymmetric, so $\hat{\bphi}\cdot \bnabla \psi = 0$, \ie $\partial_\phi \psi = 0$.  Then from $\partial_\phi (\hat{\bphi}\cdot\B_{\mathrm{cov}} )= 0$ we obtain

\begin{equation}
G \frac{\partial ^2 \varphi}{\partial \phi^2} + I \frac{\partial^2 \theta}{\partial \phi^2} = 0.
\end{equation}  
Integrating, using $\varphi = \phi + \nu$, and  periodicity in $\phi$ we obtain

\begin{equation}
G\frac{\partial \nu}{\partial \phi} + I \frac{\partial \theta}{\partial \phi} = 0.\label{axi-eqn-1}
\end{equation}
Now, taking the $\partial_\phi (\hat{\bf R}\cdot {\bf B}_{\mathrm{con}} )= 0$, we likewise obtain

\begin{equation}
\frac{\partial \theta}{\partial \phi} - \iotaslash \frac{\partial \nu}{\partial \phi} = 0.\label{axi-eqn-2}
\end{equation}
Eqns.~\ref{axi-eqn-1}-\ref{axi-eqn-2} are linearly independent (\ie $G + \iotaslash I \neq 0$ by Eqn.~\ref{JacB-eqn-1}), so we have

\begin{equation}
\frac{\partial \theta}{\partial \phi} = \frac{\partial \nu}{\partial \phi} = 0.\label{axi-eqn-3}
\end{equation}


For obtaining the Grad Shafranov equation, it is convenient to use a mixed form of $\B$, using the toroidal part of the covariant field, and the non-toroidal part of the contravariant field:

\begin{equation}
\B_{\mathrm{mix}} = G \bnabla \phi + \bnabla \phi \times \bnabla \Psi,
\end{equation}
where $\Psi(\psi)$ is the poloidal magnetic flux, and $d\Psi/d\psi = \iotaslash$.  Using this form, force balance and Ampere's law imply $\mu_0\bnabla \Psi dp/d\Psi = (\bnabla \times \B_{\mathrm{mix}}) \times \B_{\mathrm{mix}}$, which immediately yields the Grad-Shafranov equation,

\begin{equation}
\mu_0\frac{dp}{d\Psi} = -\frac{1}{R^2}\left[G\frac{dG}{d\Psi} + \Delta^* \Psi\right]
\end{equation}
where 

\begin{equation}
\Delta^* \Psi = R \frac{\partial }{\partial R}\left(\frac{1}{R}\frac{\partial \Psi}{\partial R}\right) + \frac{\partial^2 \Psi}{\partial Z^2}
\end{equation}
Although this is a complete specification of the axisymmetric field, we require the full set of coordinates to solve our the perturbative problem, so we must develop the more general representations, \ie $\B_{\mathrm{cov}}$ and $\B_{\mathrm{con}}$.  The functions $\theta(R, Z)$ and $\nu(R,Z)$ (and $I$ and $\iotaslash$ as functions of $\Psi$) can be obtained from additional equations derived from components of the MHD constraint, $\B_{\mathrm{con}} = \B_{\mathrm{cov}}$:

\begin{equation}
G(\psi) \bnabla\varphi + I(\psi) \bnabla\theta + K(\psi, \theta, \varphi) \bnabla \psi= \bnabla\psi \times \bnabla\theta - \iotaslash \bnabla \psi \times \bnabla \varphi,\label{MHD-direct-eqn}
\end{equation}

The $\hat{\bphi}$ component gives

\begin{equation}
\frac{\iotaslash G}{R} = \poiss{\theta - \iotaslash \nu}{\Psi},\label{D9}\\
\end{equation}
where $\poiss{A}{B} = \partial_R A \partial_Z B - \partial_R B \partial_Z A$, and the $\hat{\bf R}$ and $\hat{\bf Z}$ components can be combined to eliminate $K$, yielding

\begin{equation}
\frac{1}{R} |\bnabla\Psi|^2 =  \poiss{I \theta + G \nu}{\Psi}.\label{D10}
\end{equation}
Then $\psi$ can be obtained from

\begin{equation}
\frac{d\psi}{d\Psi} = \frac{1}{\iotaslash}.\label{psi-Psi-eqn}
\end{equation}
Finally, $K$ can be obtained from the $\bnabla \psi$ component of Eqn.~\ref{MHD-direct-eqn} (\ie the condition that $\B_{\mathrm{cov}}$ has no component pointing out of the magnetic surface),

\begin{equation}
K(R,Z) = \frac{1}{|\bnabla\psi|^2} ( G\bnabla\psi\cdot\bnabla\nu + I \bnabla\psi\cdot\bnabla\theta ).
\end{equation}

\subsection{Solving Eqns.~\ref{D9}-\ref{D10}}

Now we would like to solve these equations for the unknowns $\nu$ and $\theta$, and $K$.  The Boozer angle $\theta$ can be expressed as

\begin{equation}
\theta = \vartheta + \lambda(R, Z),
\end{equation}
in terms of the geometric poloidal angle $\vartheta$, defined in terms of the quadrant-specific $\arctan$ function as $\vartheta = \arctan(R-R_a, z - z_a)$, with $R_a$ and $z_a$ the coordinates of the magnetic axis.  Defining the potentials $P =  G \nu + I \lambda$ and $A = \lambda - \iotaslash \nu$, we can obtain from Eqn.~\ref{D9} the equation

\begin{align}
\frac{\partial A}{\partial \vartheta} = -1 + \frac{\iotaslash G}{R {\cal J}(\vartheta)},\\
\frac{\partial P}{\partial \vartheta} = -I + \frac{|\bnabla \Psi|^2}{R {\cal J}(\vartheta)},
\end{align}
where ${\cal J}(\vartheta) = -\phih \cdot (\bnabla \vartheta\times\bnabla\Psi) = \poiss{\vartheta}{\Psi}$.  Note that, formally, we are changing coordinates to $\vartheta$ and $\Psi$.  The functions $\iotaslash$ and $I$ may be found as solubility constraints of these two equations:

\begin{eqnarray}
&\iotaslash &= 2\pi \left( \int_0^{2\pi} d\vartheta \frac{G}{R {\cal J}(\vartheta)} \right)^{-1},\\
&I &= \frac{1}{2\pi}\int_0^{2\pi} d\vartheta \frac{|\bnabla \Psi|^2}{R {\cal J}(\vartheta)},
\end{eqnarray}
and the solutions for $A$ and $P$ given as
\begin{eqnarray}
A =  \int_0^{\vartheta} d\vartheta^\prime \left( \frac{\iotaslash G}{R {\cal J}(\vartheta^\prime)} - 1 \right),\\
P =  \int_0^{\vartheta} d\vartheta^\prime \left( \frac{|\bnabla \Psi|^2}{R {\cal J}(\vartheta^\prime)} - I\right),
\end{eqnarray}
where note that $R$ and $|\bnabla\Psi|^2$ are also evaluated at $\vartheta^\prime$, and the choice $P=A=0$ at $\vartheta=0$ has been made.  From these solutions, the desired potentials $\lambda$ and $\nu$ may be obtained as
\begin{align}
\lambda &= \frac{\iotaslash P + G A}{\iotaslash I + G},\\
\nu &= \frac{P - I A}{\iotaslash I + G},
\end{align}
where we again note that $\iotaslash I + G \neq 0$ by Eqn.~\ref{JacB-eqn-1}.  Finally, to obtain $\iotaslash$, $I$ and $G$ as functions of $\rho$, we solve Eqn.~\ref{psi-Psi-eqn}, and the functions $R(\rho, \theta)$ and $z(\rho, \theta)$, are obtained by inverting $\theta(R, z)$ and $\Psi(R,z)$.  

Restoring subscripts and normalizations, we are thus furnished with the functions $R_0(\rho, \theta)$, $z_0(\rho, \theta)$, $\nu_0(\rho, \theta)$, $\Kb_0(\rho, \theta)$, $\iotaslash_0(\rho)$, $\Ib(\rho)$, and $\Gb(\rho)$ to substitute into the expressions provided in section \ref{order-zero-sec}, and proceed to the calculations, at next order, in section \ref{ep1-sec}.

\end{document}